\documentclass[11pt,a4paper]{article}
\usepackage[utf8]{inputenc}
\usepackage{amsmath}
\usepackage{amsfonts}
\usepackage{amssymb}
\usepackage{graphicx}
\usepackage{bm}
\usepackage{booktabs}       
\usepackage{multirow}
\usepackage{epstopdf}
\usepackage{hyperref}       
\usepackage{url}            
\usepackage[top=2.52cm, bottom=2.52cm, left=2.52cm, right=2.52cm]{geometry}
\usepackage{xcolor} 
\usepackage{setspace}
\begin{document}
\title{\textbf{On the nonreciprocal propagation of light waves in asymmetric lossy system of thin films}}



\author{Dheeraj Pratap\footnote{Email: dpratap@iitd.ac.in}\\ \\
	Optics and Photonic Centre, Indian Institute of Technology Delhi, Delhi 110016, India\\
	}

\date{}

\maketitle

\begin{abstract}
Optical nonreciprocity is the phenomenon where light behaves differently when traveling in the forward direction compared to the backward direction. This nonreciprocal behavior is typically achieved in systems that exhibit electromagnetic coupling, temporal variation, and nonlinear properties. In this work, we demonstrate that nonreciprocal behavior can also exist in an optically asymmetric lossy system without requiring these conditions. In our study, we theoretically investigate the nonreciprocal propagation of light waves in a passive asymmetric optical system made from a thin film that experiences some losses. Our theoretical analysis has been experimentally validated. This type of system can be utilized in passive nonreciprocal devices.

\vspace{3mm}
\end{abstract}


\section{Introduction}
Optical nonreciprocity is a phenomenon that violates the Lorentz reciprocity theorem and prevents a light field from going back along its initial course after traveling through an optical system in a single direction. It is vital for both basic research and practical sciences~\cite{mackay2010electromagnetic}. When the positions of sources and detectors of a nonreciprocal system are switched, the received-transmitted field ratios generally shift~\cite{asadchy2020tutorial}. Nonreciprocity can be used to achieve one-way light propagation, as in optical circulators\cite{kamal2011noiseless}, isolators~\cite{jalas2013and}, and directional amplifiers~\cite{malz2018quantum}. Refraction, diffraction, mode conversion, and polarization conversion are the optical processes most closely adhering to reciprocity. Maxwell's equations' time-reversal symmetry and linear character constrain the fundamental ideas needed to realize optical nonreciprocity. Materials with asymmetric permittivity or permeability tensors, time-varying optical systems, and nonlinear light–matter interactions are the three plausible approaches for violating optical reciprocity~\cite{caloz2018electromagnetic}. The prevailing methodology relies on materials with asymmetric tensors that have the magnetic properties~\cite{bahari2017nonreciprocal}. In this method, ferrite-based technologies depend on relatively large permanent magnets and are incompatible with nanotechnology. The second method, which is based on time-varying systems, has made it possible to reduce the size of nonreciprocal components to the micron-scale. However, its weak response, power inefficiency, and general complexity of the modulation needed to function in the optical spectral domain present significant technological obstacles to further reduction to the nano-scale~\cite{ji2022compact}. Nonlinear light-matter interactions are the third method, which is the most practical technique to achieve nonreciprocity at the nanoscale. The incapacity to function under two or more simultaneous excitations is one of the main limits of nonlinearity-induced nonreciprocity. There may be a trade-off between the range of operating powers and insertion loss for nonlinear nonreciprocity, which only occurs within a specific range of incident powers~\cite{shi2015limitations}. In this regard, thermally induced phase change materials were also employed to achieve the nonreciprocity~\cite{tripathi2024nanoscale}. However, the temperature also worked as an external bias. In all three of the former approaches to achieve nonreciprocity, the optical system must have either electromagnetic coupling or biasing that works as an active asymmetric perturbation in time.  

In this work, we present a passive, lossy, and optically asymmetric system that shows the nonreciprocity in the reflection and transmission of the light intensity after its interaction. The loss and asymmetry must be present to achieve the optical nonreciprocity in our theoretically presented system. Our analysis is based on the Fresnel coefficient approach of classical optics. This type of system can be achieved easily. We verify experimentally the theoretically presented nonreciprocity of the optical asymmetric lossy system.  

\section{Theory}
\begin{figure}
\centering
\includegraphics[width=0.5\textwidth]{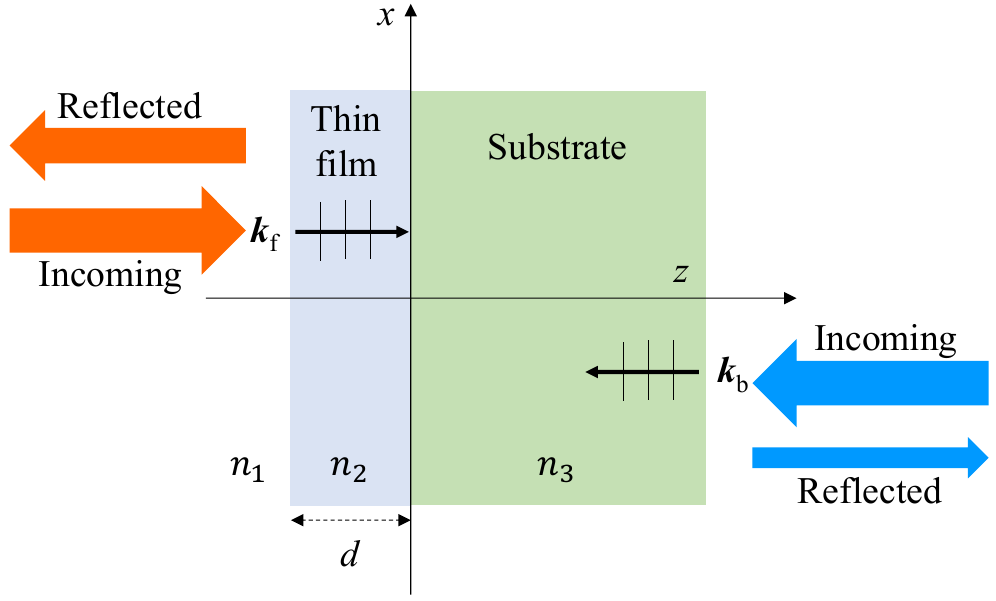}
\caption{Schematic showing the nonreciprocal reflections from the lossy asymmetric optical system of a thin film in between two mediums for the forward and backward illuminations.}
\label{fig:schematic}
\end{figure}

Consider an optically thin film of thickness $d$ that separates two media on its left and right sides. The refractive indices of the left side medium (first), thin film (second), and right side medium (third) are $n_1=n'_1+i n''_1$, $n_2=n'_2+i n''_2$, and $n_3=n'_3+i n''_3$, respectively. The propagation constants in respective mediums are $k_1$, $k_2$, and $k_3$ . The single prime and double prime represent the real and imaginary parts of the complex refractive indices of the mediums. This is an optically asymmetric system along a direction normal to the interface of the film and medium. Figure~\ref{fig:schematic} depicts a typical lossy and optically asymmetric system along the $z$-direction. The interfaces between the thin film and the mediums are in the $xy-$plane. 
At the normal incidence, the reflection and transmission coefficients at an interface for the wave propagation from $i^{\mathrm{th}}$ towards $j^{\mathrm{th}}$ medium are given by,
\begin{equation}
r_\mathrm{ij}=\frac{n_\mathrm{j}-n_\mathrm{i}}{n_\mathrm{i}+n_\mathrm{j}}, \quad
t_\mathrm{ij}=\frac{2 n_\mathrm{i}}{n_\mathrm{i}+n_\mathrm{j}}, \quad
r_\mathrm{ji}=-r_\mathrm{ij}, \quad
t_\mathrm{ij}r_\mathrm{ji}=1-r^2_\mathrm{ij}, \quad (\mathrm{i,j}=1,2,3).
\label{eq:rij_tij}
\end{equation}
For the forward (along $+z$) direction propagation, the reflection and transmission coefficients from the thin film are given by~\cite{born2013principles}, 
\begin{equation}
r_\mathrm{f}=\frac{r_{12}+r_{23}e^{2ik_2d}}{1+r_{12}r_{23}e^{2ik_2d}}, \quad
t_\mathrm{f}=\frac{t_{12}t_{23}e^{2ik_2d}}{1+r_{12}r_{23}e^{2ik_2d}}.
\label{eq:rf_tf}
\end{equation}
%
%
The reflectance and transmittance from the film of thickness $d$ are given by,
\begin{align}
R_\mathrm{f} & =\frac{r_{12} r^*_{12} + r_{12} r^*_{23}e^{-2ik_2d} + r^*_{12} r_{23}e^{2ik_2d} + r_{23} r^*_{23}}{1 + r^*_{12} r^*_{23}e^{-2ik_2d} + r_{12} r_{23}e^{2ik_2d} + r_{12} r^*_{12} r_{23} r^*_{23}}, \nonumber \\
T_\mathrm{f} & =\frac{n_\mathrm{3}}{n_\mathrm{1}} \left( \frac{t_{12} t^*_{12} t_{23} t^*_{23}}{1 + r^*_{12} r^*_{23}e^{-2ik_2d} + r_{12} r_{23}e^{2ik_2d} + r_{12} r^*_{12} r_{23} r^*_{23}} \right).
\label{eq:Rf_Tf}
\end{align}

Similarly, for the backward (along $-z$) direction propagation, the reflection and transmission coefficients are given by,
\begin{equation}
r_\mathrm{b}=\frac{r_{32}+r_{21}e^{2ik_2d}}{1+r_{32}r_{21}e^{2ik_2d}}, \quad
t_\mathrm{b}=\frac{t_{32}t_{21}e^{2ik_2d}}{1+r_{32}r_{21}e^{2ik_2d}}.
\label{eq:rb1_tb1}
\end{equation}
%
%
Using the relations of $r_\mathrm{ij}$ in Eq.~(\ref{eq:rij_tij}), the $r_\mathrm{b}$ and $t_\mathrm{b}$ can be represented in the following form,
\begin{equation}
r_\mathrm{b}=-\frac{r_{23}+r_{12}e^{2ik_2d}}{1+r_{12}r_{23}e^{2ik_2d}}, \quad
t_\mathrm{b}=\frac{t_{32}t_{21}e^{2ik_2d}}{1+r_{12}r_{23}e^{2ik_2d}}.
\label{eq:rb2_tb2}
\end{equation}
Now the reflectance and transmittance for the backward direction illumination are given by,
\begin{align}
R_\mathrm{b} & =\frac{r_{12} r^*_{12} + r^*_{12} r_{23}e^{-2ik_2d} + r_{12} r^*_{23}e^{2ik_2d} + r_{23} r^*_{23}}{1 + r^*_{12} r^*_{23}e^{-2ik_2d} + r_{12} r_{23}e^{2ik_2d} + r_{12} r^*_{12} r_{23} r^*_{23}}, \nonumber \\
T_\mathrm{b} & =\frac{n_\mathrm{1}}{n_\mathrm{3}} \left( \frac{t_{21} t^*_{21} t_{32} t^*_{32}}{1 + r^*_{12} r^*_{23}e^{-2ik_2d} + r_{12} r_{23}e^{2ik_2d} + r_{12} r^*_{12} r_{23} r^*_{23}} \right).
\label{eq:Rb_Tb}
\end{align}
For a reciprocal system, the forward and backward reflectance and or  transmittance should be equal, i.e., $R_\mathrm{f}=R_\mathrm{b}$ and or $T_\mathrm{f}=T_\mathrm{b}$. If the system is nonreciprocal then $R_\mathrm{f} \ne R_\mathrm{b}$ and or $T_\mathrm{f} \ne T_\mathrm{b}$. Let's define two functions, 
\begin{align}
F^\mathrm{}_R (n_1,n_2,n_3)= &R_\mathrm{12} R^*_\mathrm{23}-R^*_\mathrm{12} R_\mathrm{23} \nonumber \\
= & \frac{2(n^2_2 n^*_2 n^*_3 - n_2 {n^*_2}^2 n_3 + n^*_1 n_2 n_3 n^*_3 - n^*_1 n^2_2 n^*_2 - n_1 n^*_2 n_3 n^*_3 + n_1 n_2 {n^*_2}^2 -n_1 n^*_1 n_2 n^*_3 + n_1 n^*_1 n^*_2 n_3)}{(n_1+n_2)(n^*_1+n^*_2)(n_2+n_3)(n^*_2+n^*_3)},
\label{eq:FR}
\end{align}
and 
\begin{align}
F^\mathrm{}_T (n_1,n_2,n_3) = & \frac{n_3}{n_1} \left(T_{12} T^*_{12} T_{23} T^*_{23}\right) - \frac{n_1}{n_3} \left(T_{21} T^*_{21} T_{32} T^*_{32}\right) \nonumber \\
= & \frac{16 n_2 n^*_2 (n^*_1 n_3 - n_1 n^*_3)}{(n_1+n_2)(n^*_1+n^*_2)(n_2+n_3)(n^*_2+n^*_3)}.
\label{eq:FT}
\end{align}

For the nonreciprocal system, either function $F^\mathrm{}_R (n_1,n_2,n_3)$ or $F^\mathrm{}_T (n_1,n_2,n_3)$, or both must have non-zero values. From Eqs.~(\ref{eq:FR}) and (\ref{eq:FT}), it is clear that in general, the complex refractive index as well as an optically asymmetric system is nonreciprocal for reflectance and transmittance. Usually, this type of nonreciprocal effect is quite small for a low-loss system, but it can be increased by choosing a suitable combination. To make a thin film system nonreciprocal, the function $F^\mathrm{}_R $ and $F^\mathrm{}_T $ must be non-zero at some values of refractive indices $n_1,n_2$, and $n_3$. Let's consider two simpler cases to analyze this type of nonreciprocal effect. 

Case I: If the first and third mediums are lossless ($n''_1=n''_3=0$), and the second medium is lossy ($n''_2\ne 0$), then the function $F^\mathrm{}_R (n_1,n_2,n_3)$ can be written in the following form,
\begin{equation}
F^\mathrm{I}_R (n'_1,n'_2+i n''_2,n'_3)= \frac{2(n'_3-n'_1)(n_2-n^*_2)(n_2 n^*_2+n'_1 n'_3)}{(n'_1+n_2)(n'_1+n^*_2)(n_2+n'_3)(n^*_2+n'_3)},
\label{eq:F1R}
\end{equation}
and
\begin{equation}
F^\mathrm{I}_T (n'_1,n'_2+i n''_2,n'_3) = 0.
\label{eq:F1T}
\end{equation}

Case II: If the first medium and thin film are lossless ($n''_1=n''_2=0$) and the third medium is lossy ($n''_3\ne 0$), then the function $F^\mathrm{}_R (n_1,n_2,n_3)$ becomes,
\begin{equation}
F^\mathrm{II}_R (n'_1,n'_2,n'_3+i n''_3)=\frac{2 n'_2 (n'_1-n'_2)(n'_1+n'_2) (n_3-n^*_3)}{(n'_1+n'_2)^2 (n'_2+n_3)(n'_2+n^*_3)},
\label{eq:F2R}
\end{equation}
and 
\begin{equation}
F^\mathrm{II}_T (n'_1,n'_2,n'_3+i n''_3) = \frac{16 n'^2_2 n'_1 (n_3 - n^*_3)}{(n'_1+n'_2)^2 (n'_2+n_3)(n'_2+n^*_3)}.
\label{eq:F2T}
\end{equation}
From Eqs.~(\ref{eq:F1R}), (\ref{eq:F1T}), (\ref{eq:F2R}) and (\ref{eq:F2T}), we see that to become thin film system nonreciproical, it should be optically asymmetric and atleast one medium should have complex refractive index. In the optical system of Case I, reflection is nonreciprocal while transmission is reciprocal, even if the system is asymmetric. For the optical system of Case II, reflection and transmission are both nonreciprocal. However, the nonreciprocity of transmission will be extremely small or negligible due to significant loss in the thicker (semi-infinite) third medium (or equivalently, first medium). Reflection is nonreciprocally dominating and interesting than transmission, which can be easily observed in practical cases. We define $\Delta R = R_\mathrm{f}-R_\mathrm{b} $ to quantify the strength of reflection nonreciprocity. Higher values of the $\Delta R$ represent larger nonreciprocity. The positive and negative signs of $\Delta R$ will indicate domination of the reflection for wave propagation along the forward and backward directions, respectively.

\begin{figure}
\centering
\includegraphics[width=0.9\textwidth]{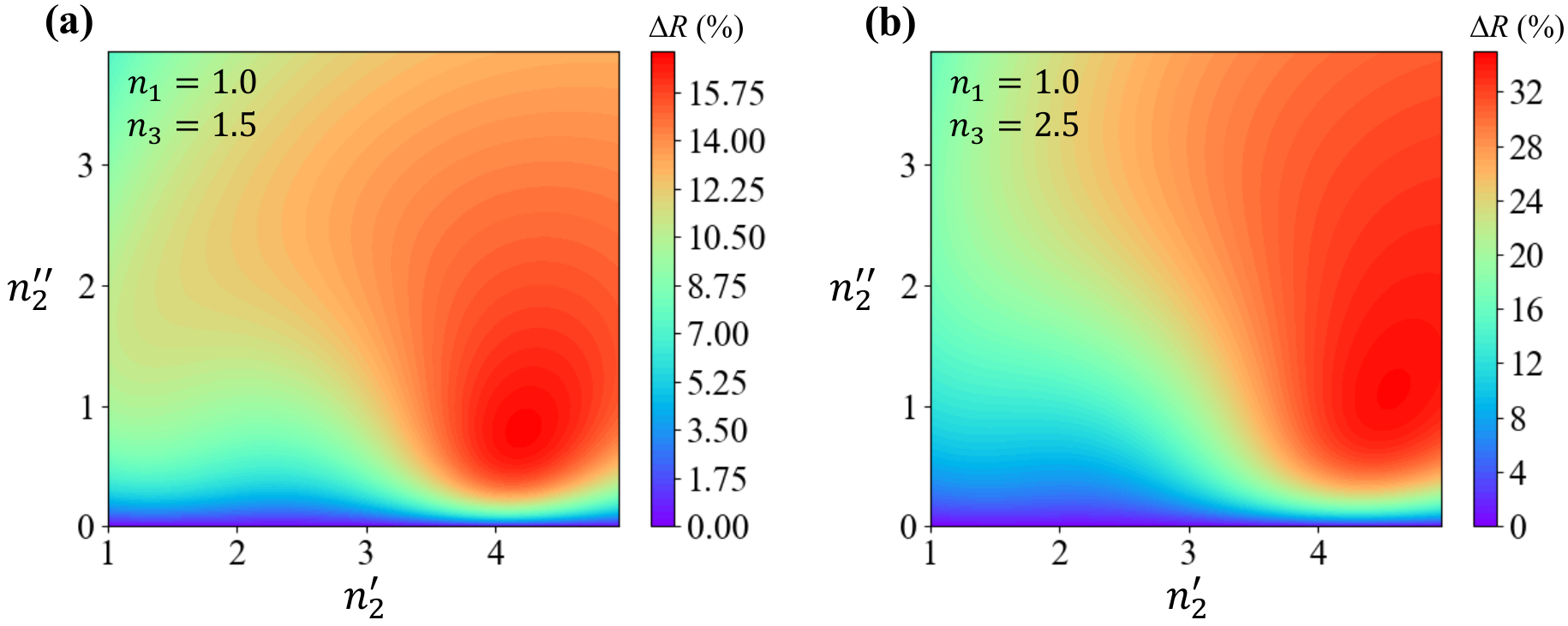}
\caption{Difference of reflectance $\Delta R$ for the wave propagation along forward and backward directions with variation of complex refractive index of thin film at (a) $n_1 = 1.0$, $n_3 = 1.5$, and (b) $n_1 = 1.0$, $n_3 = 2.5$.}
\label{fig:n2_loss_variation}
\end{figure}

\begin{figure}
\centering
\includegraphics[width=0.9\textwidth]{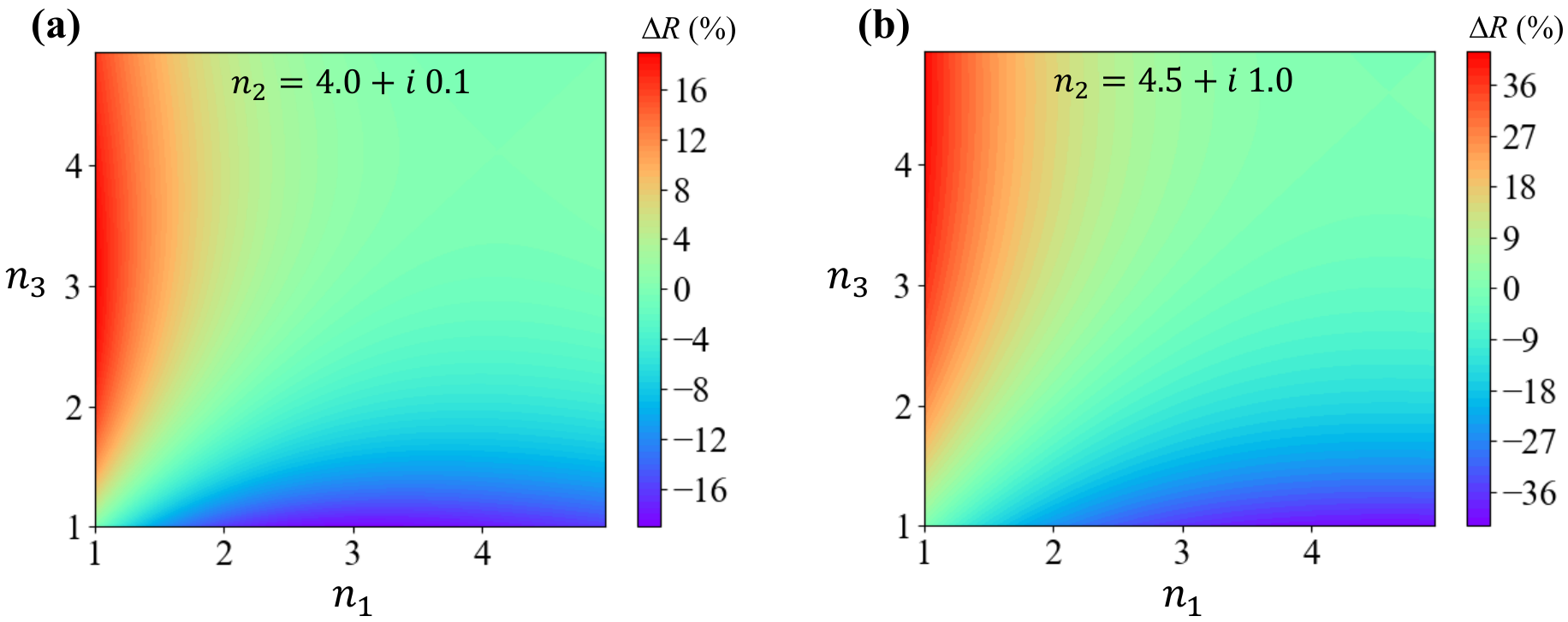}
\caption{Difference of reflectance $\Delta R$ for the wave propagation along forward and backward directions with variation of refractive indexes of first and third medium at (a) $n_2 = 4.0 + i 0.1$, and (b)  $n_2 = 4.5 + i 1.0$.}
\label{fig:n3_loss_variation}
\end{figure}

\begin{figure}
\centering
\includegraphics[width=1\textwidth]{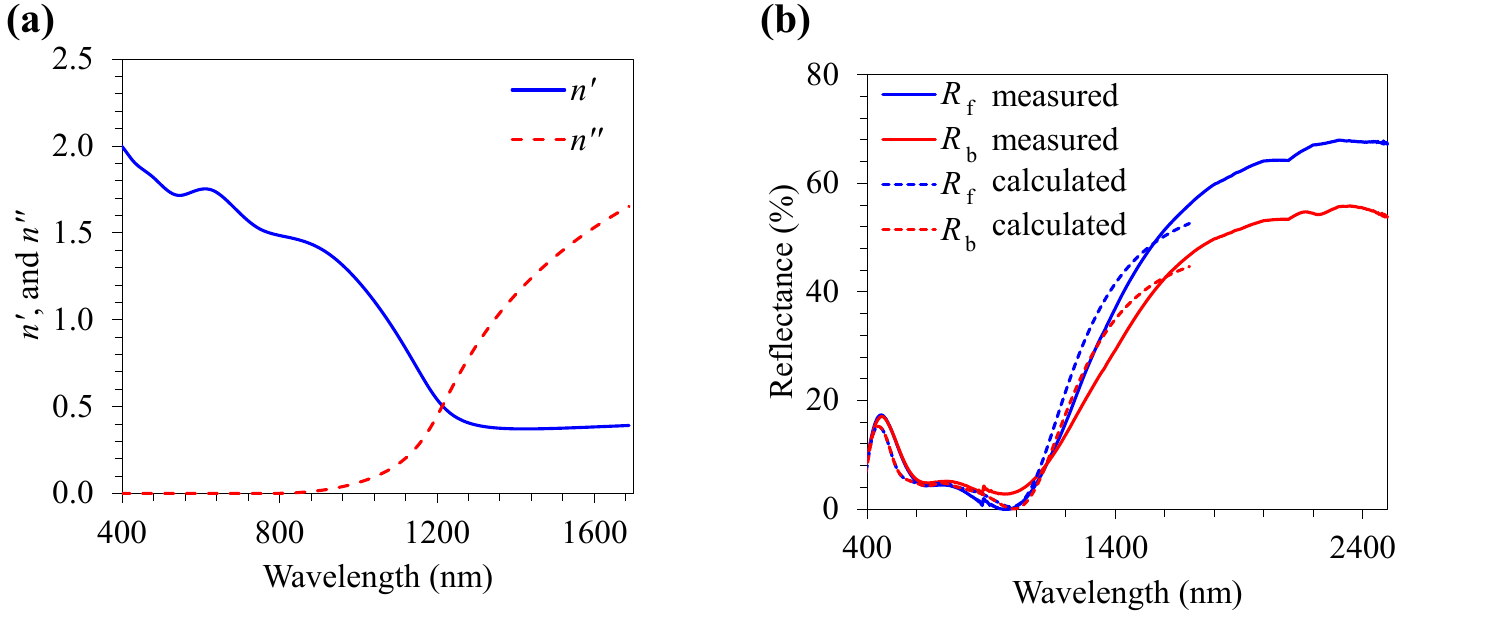}
\caption{(a) Measured complex refractive index of the Indium Tin Oxide (ITO) thin film on the Soda Lime glass substrate kept in air. (b) Measured and analytically calculated forward and backward reflectance of the optical system in (a).}
\label{fig:Rf_Rb_exp_cal}
\end{figure}

\section{Discussion}
For the practical scenario, a lossy thin film on a substrate in air for Case~I, or a lossless thin film on a lossy substrate in air for Case~II, can work as an asymmetric optical system. Figure~\ref{fig:n2_loss_variation} shows the reflection difference $\Delta R$ for the illuminations along the forward and backward directions. In Fig.~\ref{fig:n2_loss_variation}(a), the first and third media have refractive indices of 1 and 1.5 (lossless), respectively, and the real and imaginary parts of the complex refractive index of the thin film are varied to see the $\Delta R$. For Fig.~\ref{fig:n2_loss_variation}(b), the refractive index of the third medium is 2.5 while keeping the same index as the former case. We see that the $\Delta R$ for the former case is approximately 17\% while for the latter case, 35\%. It is due to the fact that the index asymmetry of the first and third mediums increases the $\Delta R$. For example, the glass and Titanium Dioxide (TiO$_2$) can work as substrates for lower and higher optical asymmetry. It should be noted that the maximum $\Delta R$ occurs when the $n''_2$ is less than 1. Therefore, to get the higher reflection nonreciprocity, the losses may not be larger, but they should have an optimum value.

Now, we fix the value of the complex refractive index of the thin film and vary the lossless indices of the first and third mediums. Figure~\ref{fig:n3_loss_variation} shows the density plots of the reflection nonreciprocity $\Delta R$ with the variation of refractive indices of the first and third medium. In the Fig.~\ref{fig:n3_loss_variation}(a), the value of the complex refractive index is $n_2=4.0+i0.1$ (a typical value of Silicon index in near infra-red). At this complex index, the maximum $\Delta R$ is approximately 20\%. According to our convention, if the first medium index is higher than the third, then the $\Delta R$ is negative, which means more reflection will be in the third medium for the backward propagation. If the index of the first medium is lower than the third, then $\Delta R$ is positive, which means reflection will be higher in the first medium for the forward propagation. In the Fig.~\ref{fig:n3_loss_variation}(b), the complex refractive index is $n_2=4.5+i1.0$ (a typical value of Silicon index in mid visible). We see that the reflection nonreciprocity has the same trends as in the former case. However, now the $\Delta R$ has a maximum value around 45\% for the given ranges of the first and third medium indices. 

For the experimental validation of the analytical results, we took the thin film of Indium Tin Oxide (ITO) coated on the Soda Lime glass kept in air. The air was the first medium there, and the substrate was optically semi-infinite and worked as the third medium. That configuration works as an optically asymmetric lossy system of thin film. The thickness of the ITO film was 185~nm. The complex refractive index of the ITO thin film was obtained from the Ellipsometry as shown in Fig.~\ref{fig:Rf_Rb_exp_cal}(a). We can see that a significant loss starts around 1000~nm, and below that, losses are approximately zero. Figure~\ref{fig:Rf_Rb_exp_cal}(b) shows the measured and analytically calculated reflectance for the forward and backward illuminations. The Barium sulfate diffuser was used as a reference for the reflection measurement. As the loss of the ITO thin film increases,  the reflection difference increases for the illuminations from the forward and backward directions. Ellipsometry data of ITO were used for the analytical calculations of the reflectance. The calculated results show a steeper slope than the experimental. This might be due to the fact that, for the calculation, there was no need for any reference, while it was needed for the experiment. The ellipsometry data were available up to 1688~nm, while the measured reflection was up to 2500~nm. Therefore, the calculated data range was not up to the experimental reflection. Our experimental and analytical data show a good understanding of the nonreciprocity in the lossy asymmetric thin film system.

\section{Conclusion}
In summary, we presented a theory of nonreciprocity in a passive system that could be realised in the asymmetric lossy system of thin films. Such a system could be easily implemented for practical use. We demonstrated the experimental validation of the theoretical prediction with good agreement. Such a system could be utilised to design the photonic system for the nonreciprocal devices.

\section*{Acknowledgement}
D. P. thanks IIT Delhi for the NFSG fund project number MI02922G, and ANRF (SERB) for the fund project number EEQ/2023/000240. 




\bibliography{references}   
\bibliographystyle{ieeetr}
%
%
%
\end{document}